\documentclass[fleqn, twoside, 11pt]{article}
\usepackage{graphics}
\begin{document}
\title{Quasistationary collapse to the extreme Kerr black hole}
\author{Reinhard Meinel\\ University of Jena,
Institute of Theoretical Physics,\\
Max-Wien-Platz  1, 07743 Jena, Germany}
\date{meinel@tpi.uni-jena.de}
\maketitle
\thispagestyle{empty}
\begin{abstract}
It is shown that the extreme Kerr black hole is the only candidate
for a black hole limit of rotating fluid bodies in equilibrium.
\end{abstract}

\newpage
\section{Introduction}

The gravitational collapse of a spherically symmetric star to the
Schwarzschild black hole is well understood. The idealized model
of dust collapse described by Oppenheimer and Snyder \cite{os}
provides a famous analytic example. In contrast, the belief that
the collapse of a {\it rotating} star leads finally to the Kerr
black hole relies on Penrose's cosmic censorship hypothesis
combined with the black hole uniqueness theorems (see \cite{he,
heu}) and has not been rigorously proved so far. As a step in this
direction we want to discuss in this paper the possibility of a
{\it quasistationary} collapse to the Kerr black hole, i.e.\ the
possibility of sequences of equilibrium configurations of stars
approaching a black hole limit continuously. Of course, the
surface of a stationary star can never become identical with a
black hole horizon since the horizon is a null hypersurface. The
question is, however, whether it is possible to come arbitrarily
close to this limit. In the nonrotating, i.e.\ spherically
symmetric case, the answer is ``no''. Under the reasonable
assumption, that the mass-energy density is non-negative and does
not increase outwards, the circumferential radius
$r_*$ of a static fluid sphere of total mass $M$ always satisfies
the ``Buchdahl inequality'' $r_* > (9/8)\, r_s$, where $r_s=2M$ is
the corresponding Schwarzschild radius (see, for example,
\cite{wald}). Consequently, the relative redshift of photons
emitted from the surface of the star and received at infinity is
bounded by the value $2$.

In the following, we consider stationary and axisymmetric,
uniformly rotating perfect fluid configurations with a ``cold''
equation of state\footnote{The results are valid for isentropic
stellar models with non-zero temperature as well.} and their
possible connection to black holes. The relevant basic equations
can be found, for example, in \cite{hs} -- \cite{nm1}. We use
units where the speed of light $c$ as well as Newton's
gravitational constant $G$ are equal to 1 and choose the spacetime
signature ($-+++$).
\section{Equilibrium stellar models, Kerr black holes and
their possible connection}

In this chapter we shall analyse necessary parameter relations for
a possible connection between equilibrium stellar models and Kerr
black holes. Therefore we have to provide some basic relations for
rotating fluids in equilibrium.

The four-velocity of the fluid must point in the direction of a
linear combination of the two Killing vectors
$\xi=\partial/\partial t$ and $\eta=\partial/\partial \phi$
corresponding to stationarity and axisymmetry:
\begin{equation}
u^i=e^{-V}(\xi^i+\Omega\,\eta^i),\quad \Omega=\mbox{\rm constant.}
\end{equation}
The Killing vector $\xi$ is fixed by its normalization
$\xi^i\xi_i\to -1$ at spatial infinity (we assume asymptotic
flatness). The orbits of the spacelike Killing vector $\eta$ are
closed and $\eta$ is zero on the axis of symmetry.
$\Omega=u^\phi/u^t$ is the angular velocity of the fluid body with
respect to infinity. The factor $e^{-V}=u^t$ follows from
$u^iu_i=-1$:
\begin{equation}
(\xi^i+\Omega\,\eta^i)(\xi_i+\Omega\,\eta_i)=-e^{2V}.
\label{V}
\end{equation}
The energy-momentum tensor is given by
\begin{equation}
T_{ik}=(\epsilon+p)\,u_iu_k+p\,g_{ik},
\end{equation}
where the mass-energy density $\epsilon$ and the pressure $p$ are
related by a ``cold'' equation of state, $\epsilon=\epsilon(p)$,
following from \noindent
\begin{equation}
p=p(\rho,T),\quad \epsilon=\epsilon(\rho,T),\quad  \mbox{$\rho$:
baryonic mass-density}
\end{equation}
for $T=0$. The specific enthalpy
\begin{equation}
h=\frac{\epsilon+p}{\rho}
\label{h}
\end{equation}
can be calculated as follows:
\begin{equation}
dh=Tds+\frac{1}{\rho}\,dp=\frac{1}{\rho}\,dp, \quad \mbox{$s$:
specific entropy}
\end{equation}
\begin{equation} \Rightarrow \frac{dh}{h}=\frac{dp}{\epsilon+p} \Rightarrow
h(p)=h(0)\,\exp\left[\int\limits_0^p\frac{dp'}{\epsilon(p')+p'}\right].
\end{equation}
Note that $h(0)=1$ for most equations of state. For our
conclusions, however, it will be sufficient to assume
$h(0)<\infty$. From ${T^{ik}}_{;k}=0$ we obtain
\begin{equation}
h(p)\,e^V=h(0)\,e^{V_0}={\rm constant},
\label{cc}
\end{equation}
where $V_0$, the constant surface value (corresponding to $p=0$)
of the function $V$ defined in (\ref{V}), is related to the
relative redshift $z$ of zero angular momentum
photons\footnote{This means $\eta_ip^i=0$ ($p^i$: four-momentum of
the photon).} emitted from the surface of the fluid and received
at infinity:
\begin{equation}
z=e^{-V_0}-1.
\end{equation}
Equilibrium models, for a given equation of state, are fixed by
two parameters, for example $\Omega$ and $V_0$. The total mass $M$
and angular momentum $J$ are given by
\begin{equation}
M=2\int\limits_{\Sigma}(T_{ik} - \frac{1}{2}Tg_{ik})n^i\xi^kd
{\cal V}, \quad J=-\int\limits_{\Sigma}T_{ik}\,n^i\eta^kd{\cal V},
\label{MJ}
\end{equation}
where $\Sigma$ is a spacelike hypersurface ($t={\rm constant}$)
with the volume element $d{\cal V}=\sqrt{^{(3)}g}\,d^3x$ and the
future pointing unit normal $n^i$, see \cite{wald}. The baryonic
mass $M_0$ corresponding to the local conservation law $(\rho
u^i)_{;i}=0$ is
\begin{equation}
M_0=-\int\limits_{\Sigma}\rho\,u_i\,n^id{\cal V}.
\label{M0}
\end{equation}
Note that nearby equilibrium configurations with the same equation
of state are related by \cite{hs, bw}
\begin{equation}
\delta M=\Omega\, \delta J + h(0)\,e^{V_0}\,\delta M_0.
\end{equation}
In contrast to the non-rotating case, where Birkhoff's theorem
implies that the exterior spacetime is always given by the
Schwarzschild metric, the exterior metric of a rotating star is
not Kerr in general. However, we now want to ask whether it is
possible to approach a situation where the surface of the star
becomes ``almost'' the horizon of a Kerr black hole. First we note
that the black hole horizon is characterized by a linear
combination $\xi + \Omega^H \eta$ of the two Killing vectors which
is null \cite{ca}, where $\Omega^H$ is the ``angular velocity of
the horizon'':
\begin{equation}
(\xi^i+\Omega^H\eta^i)(\xi_i+\Omega^H\eta_i)=0 \quad \mbox{with}
\end{equation}
\begin{equation}
\Omega^H= \frac{J}{2M^2\left[M+\sqrt{M^2-(J/M)^2}\,\right]}.
\label{Om1}
\end{equation}
Since all other linear combinations  $\xi + \Omega\,\eta$
of the Killing vectors become
spacelike on the horizon the desired black hole limit can only be
approached for
\begin{equation}
\Omega=\Omega^H.
\label{Om2}
\end{equation}
Because of (\ref{V}) and (\ref{cc}), the surface of the fluid is
characterized in general by
\begin{equation}
(\xi^i+\Omega\,\eta^i)(\xi_i+\Omega\,\eta_i)=-e^{2V_0}.
\end{equation}
A black hole limit therefore requires
\begin{equation}
V_0\to -\infty \quad (z\to\infty).
\end{equation}
A combination of (\ref{MJ}) and (\ref{M0}) leads to the formula
\begin{equation}
M=2\Omega J+\int\frac{\epsilon+3p}{\rho}\,e^VdM_0,
\end{equation}
cf.\ equation (II.28) in \cite{bw}. With (\ref{h}) and (\ref{cc})
we get finally
\begin{equation}
M=2\Omega J+
h(0)\,e^{V_0}\int\frac{\epsilon+3p}{\epsilon+p}\,dM_0.
\end{equation}
Since $(\epsilon+3p)/(\epsilon+p)\le 3$ (the weak energy condition
is sufficient for this) the second term on the right-hand side of
the last equation is bounded by $3h(0)\,e^{V_0}M_0$. Consequently,
for finite $h(0)$ and finite baryonic mass $M_0$, the limit
$V_0\to -\infty$ implies
\begin{equation}
M=2\Omega J.
\end{equation}
Together with (\ref{Om1}) and (\ref{Om2}) this leads to our
conclusion that the only possible candidate for a black hole limit
of fluid bodies in equilibrium is the ``extreme'' Kerr black hole
characterized by the maximal possible value of $|J|/M^2$, i.e.\
\begin{equation}
J=\pm M^2.
\end{equation}
\section{Discussion}
Now that we have shown that the extreme Kerr black hole is the only
candidate, we want to discuss if and how this limit can be reached. In
the case of a rotating, infinitesimally thin disc, strong
numerical evidence for this transition was
found by Bardeen and Wagoner \cite{bw} who were also able to establish
its nature. Neugebauer and Meinel \cite{nm2, nm3},
see also \cite{m},
confirmed these results in finding and analysing
the exact solution to the disc problem. Hence, the existence
of a continuous connection of equilibrium fluid configurations to the extreme Kerr black
hole has been proved. (Note that the disc is a limiting case of a rotating
fluid for $p/\epsilon\to 0$.)

The horizon of the {\it extreme} Kerr black hole has
an infinite proper distance from any place in the exterior region
characterized, in Boyer-Lindquist coordinates, by $r>M$. Indeed,
the $r>M$ part of the extreme Kerr metric results from the limit
$V_0\to -\infty$ of the disc solution in the ``exterior'',
asymptotically flat region. A different limit is obtained,
however, if the limit is taken in rescaled and corotating
coordinates corresponding to $r=M$. The resulting ``inner world''
contains the disc and is not asymptotically flat but approaches
the ``extreme Kerr throat geometry'' \cite{bh} at spatial
infinity. Strictly speaking, there is not yet a horizon. The
``inner'' and the ``outer'' world are separated by the ``throat region'',
which is also characterized by $r=M$. In this subtle way the black
hole limit is realized. Presumably, a small perturbation will
lead to a genuine Kerr black hole with $|J|$ slightly
less than $M^2$.

It should be possible to prove that the above scenario holds
whenever the limit $V_0\to -\infty$ of a rotating fluid body is
attainable. A further example has already been given by the
``relativistic Dyson rings'' \cite{akm}, see also \cite{a+}.
Other ring solutions with a variety of equations of state also
exhibit this limit \cite{tf, sh}.

A perturbation analysis based on this quasistationary, ``maximally
rotating'' route to the Kerr black hole may lead to some insight
into the dynamical collapse of rotating stars supplementary to the
investigations close to the non-rotating case.

I gratefully acknowledge the many inspiring discussions with G.\
Neugebauer, A.\ Kleinw\"achter, M.\ Ansorg and D.\ Petroff.

\end{document}